\documentclass[aps,prd,twocolumn,groupedaddress]{revtex4-1}
\usepackage{graphicx}
\include{amssym}

\begin{document}
\title{Mending $\rho\pi\pi$ vertex through the $\pi a_1$ diagonalization}
\author{A. A. Osipov\footnote{Email address: osipov@nu.jinr.ru}}
\affiliation{Joint Institute for Nuclear Research, Bogoliubov Laboratory of Theoretical Physics, 141980 Dubna, Russia}

\begin{abstract}
The problem of the strong momentum dependence of the $\rho\pi\pi$-vertex in the extended Nambu -- Jona-Lasinio model is solved by an appropriate choice of fields for spin-1 particles. A corresponding phenomenological Lagrangian is derived. As straightforward applications, the decay widths of $\rho\to\pi\pi$, and $a_1\to\rho\pi$ transitions are calculated and compared with known empirical data and previous theoretical estimates.
\end{abstract}


\maketitle

\section{Introduction}
The $\pi a_1$ diagonalization is a standard procedure in the majority of the effective meson Lagrangians with spin-0 and spin-1 states. It is generally accepted that such local chiral Lagrangians arise as a result of the underlying quark-gluon dynamics \cite{Hooft74a,Hooft74b,Witten79} in the large $N_c$ limit of QCD, where $N_c$ is the number of colors. They can also be constructed on pure symmetry grounds by using different frameworks \cite{Weinberg68,Coleman69a,Coleman69b,Gasiorovicz69,Schechter84a,Schechter84b,Schechter85,Volkov86,Ebert86,Meissner88,Bando88,Bando88b,Ecker89}. Through the $\pi a_1$ diagonalization one introduces the real physical axial-vector field $a'_{1\mu} $, corresponding to the $I^G(J^{PC})=1^-(1^{++}$) $a_1(1260)$-meson, and fixes its chiral transformation laws. The simplest replacement can be written as $a_{1\mu}=a_{1\mu}'+k\partial_\mu\pi$, where $k$ is a constant, and $\pi$ is a pion field. Our recent studies \cite{Osipov17jetpL,Osipov17annals,Morais17} have shown how its explicit form influences the transformation laws of spin-1 fields.

There is a widely known difficulty with this replacement: upon proper diagonalization so-called non-minimal (higher derivative) terms enter the $\rho\pi\pi$ vertex generating an unwanted strong momentum dependence of the $f_{\rho\pi\pi}(l^2,q_1^2,q_2^2)$ coupling, i.e. $f_{\rho\pi\pi}(m_\rho^2,0,0)=\frac{3}{4} f_{\rho\pi\pi}(0,0,0)$ \cite{Schwinger67}, where $l, q_1, q_2$ are the 4-momenta of the $\rho$-meson and of the two pions. This, of course, disagrees strongly with the successful ideas of $\rho$-universality, KSRF relations, and with the experimental data. To resolve the problem one usually adds new higher derivative terms to the effective meson Lagrangian so as to cancel the unwanted strong momentum dependence of $f_{\rho\pi\pi}$ \cite{Meissner88,Bando88b}.

Although this is most probably the way out of the problem, the scheme is not applicable to the NJL model. The reason is that in the NJL model the meson vertices come out of explicit calculations of one-quark loops \cite{Volkov86}, or equivalently, from the heat kernel expansion of the chiral quark determinant \cite{Ebert86}. Thus, there is no way to insert some non-minimal meson vertices "by hand". On the other hand, it is known that all chiral Lagrangian approaches aimed to incorporate the spin-1 mesons are in principle equivalent \cite{Birse96}. Each corresponds to a different choice of fields for spin-1 mesons. It indicates that one should expect a solution which can equally well work for all approaches without an exclusion.

What mechanism might be relevant to this solution? The purpose of this paper is to show that a simple modification of the $\pi a_1$ diagonalization procedure may resolve the problem of the strong momentum dependence of the $\rho\pi\pi$ vertex in the NJL model restoring the $\rho$-universality. We also argue that this scheme can be easily extended beyond the framework of the NJL model.

\section{Effective Lagrangian}
\label{Lag}
Our starting point is the bosonized version of the Nambu--Jona-Lasinio model with $SU(2)\times SU(2)$ chiral symmetric four-quark interactions. The Lagrangian density
\begin{eqnarray}
\label{lag}
&&{\cal L}=\bar q(i\gamma^\mu\partial_\mu -{\cal M})q + {\cal L}_{S} +{\cal L}_{V}, \\
\label{lagsp}
&&{\cal L}_{S}=\frac{G_S}{2}\left[(\bar qq)^2+(\bar qi\gamma_5\vec\tau q)^2 \right], \\
\label{lagva}
&&{\cal L}_{V}=-\frac{G_V}{2}\left[(\bar q\gamma^\mu\vec\tau q)^2+(\bar q\gamma^\mu\gamma_5\vec\tau q)^2 \right]
\end{eqnarray}
includes spin-0, $G_S$, and spin-1, $G_V$, four-quark couplings; ${\cal M}=\hat m\tau_0$, $\hat m = \hat m_u=\hat m_d$ are the current quark masses (the isospin symmetry is assumed); $\tau_0$ is a unit $2\times 2$ matrix, $\vec\tau$ are the $SU(2)$ Pauli matrices; $\gamma^\mu$ are the standard Dirac matrices in four dimensional Minkowski space; in the notation of the quark field $q$ the color, isospin and 4-spinor indices are suppressed.

To describe correctly the low-energy limit, the Lagrangian must be bosonized and the terms with the smallest possible number of derivatives selected. The heat kernel technique adjusts the derivative expansion in such a way that chiral symmetry is protected.  We refer to
\cite{Osipov17annals} for a complete and exhaustive treatment of the model. For our purpose here we need the following result of such calculations
\begin{eqnarray}
\label{linv}
\mathcal L_{B}&=&-\frac{\hat m}{4mG_S}\,\mbox{tr}\, (s^2+p^{2})
+\frac{1}{4G_V}\mbox{tr}\,\left(v_\mu^{\,2}+a_\mu^{\,2}\right) \nonumber \\
&+&\frac{N_cJ_1}{16\pi^2}\mbox{tr}\left\{(\bigtriangledown_\mu s )^2 + (\bigtriangledown_\mu p)^2 \right. \nonumber \\
&-&\left. (s^2-2ms +p^2)^2 -\frac{1}{3}(v_{\mu\nu}^2+a_{\mu\nu}^2) \right\},
\end{eqnarray}
where trace is taken over isospin matrices, the covariant derivatives of the scalar $s$ and pseudoscalar $p=\vec p\vec \tau , \vec p=(p_1,p_2,p_3)$ fields together with the strengths of vector $v_\mu =\vec v_\mu\vec\tau$ and axial-vector $a_\mu =\vec a_\mu\vec\tau$ fields are given by
\begin{eqnarray}
\label{covder}
\bigtriangledown_\mu s &=&\partial_\mu s -\{a_\mu , p \}, \nonumber \\
\bigtriangledown_\mu p&=&\partial_\mu p-i[v_\mu, p]+\{a_\mu ,  s -m\}, \nonumber \\
v_{\mu\nu}&=&\partial_\mu v_\nu - \partial_\nu v_\mu -i[v_\mu , v_\nu ] -i[a_\mu , a_\nu ], \nonumber \\
a_{\mu\nu}&=&\partial_\mu a_\nu - \partial_\nu a_\mu -i[v_\mu,a_\nu ] -i[a_\mu , v_\nu].
\end{eqnarray}
The constant $m$ represents the contribution to the constituent quark masses due to spontaneous chiral symmetry breaking. It is  determined by the condition which cancels the $s$-tadpole term in $\mathcal L_B$. This condition is known as the gap equation
\begin{equation}
\label{gap}
m-\hat m= mG_S \frac{N_c}{2\pi^2}J_0(m^2,\Lambda^2),
\end{equation}
where
\begin{equation}
J_0(m^2,\Lambda^2)=\Lambda^2-m^2\ln\left(1+\frac{\Lambda^2}{m^2}\right).
\end{equation}
It is assumed that the strength of the quark interactions is large enough, $G_S>2\pi^2/(N_c\Lambda^2)$, to generate a non-trivial, $m\neq 0$, solution of eq.(\ref{gap}). The non-zero value of $m$ is held to signal the condensation of quark-antiquark pairs in the vacuum, i.e. dynamical chiral symmetry breaking. A finite ultraviolet cutoff $\Lambda$ restricts the region of integration in the one-quark-loop integrals and characterizes the energy scale where the model is applicable. Eq. (\ref{linv}) contains a further remainder of such one-quark-loop integrations. It is a function $J_1(m^2)$ given by
\begin{equation}
J_1(m^2,\Lambda^2)=\ln\left(1+\frac{\Lambda^2}{m^2}\right)-\frac{\Lambda^2}{\Lambda^2+m^2}.
\end{equation}

Let us write down the chiral $SU(2)\times SU(2)$ infinitesimal transformations of meson fields in the non-symmetric phase. They are
\begin{eqnarray}
\label{vat2}
&&\delta s =\{\beta, p\}, \nonumber \\
&&\delta p=i[\alpha, p]-\{\beta,  s-m\}, \nonumber \\
&&\delta v_\mu =i[\alpha, v_\mu ]+i[\beta, a_\mu ] , \nonumber \\
&&\delta a_\mu =i[\alpha, a_\mu ]+i[\beta, v_\mu]
\end{eqnarray}
with the parameters $\alpha =\vec\alpha\vec\tau /2$ and $\beta =\vec\beta\vec\tau /2$. It follows then that
\begin{equation}
\label{sb}
\delta\mathcal L_{B} =-\frac{\hat m}{G_S}\vec\beta\vec p.
\end{equation}

To avoid the $p a_\mu$ mixing term in $\mathcal L_{B}$ one should define a new axial-vector field, $a'_\mu$, through the replacement
\begin{equation}
\label{shift1}
a_\mu =a'_\mu +\kappa m\partial_\mu p.
\end{equation}
This changes the longitudinal component of the axial-vector field. The factor $\kappa$ is fixed by the diagonalization condition
\begin{equation}
\label{kappa}
\frac{1}{2\kappa}=m^2+ \frac{\pi^2}{N_cG_VJ_1}.
\end{equation}
Due to (\ref{kappa}) the Lagrangian density $\mathcal L_{B}$ does not contain the $a'_\mu\partial^\mu p$ term. Note that (\ref{shift1}) induces the following changes in the chiral transformation law of $a'_\mu$, and, as a consequence, of its chiral partner $v_\mu$ \cite{Osipov17jetpL,Osipov17annals}
\begin{eqnarray}
\label{vat3}
&&\delta a'_\mu =i[\alpha, a'_\mu ]+i[\beta, v_\mu]+\kappa m \{\beta,\partial_\mu s\}, \nonumber \\
&&\delta v_\mu =i[\alpha, v_\mu ]+i[\beta, a'_\mu +\kappa m\partial_\mu p].
\end{eqnarray}

Since the free part of $\mathcal L_{B}$ must preserve its canonical form, one must redefine the fields as
\begin{equation}
\label{ren}
s=g_\sigma \sigma, \ \vec p=g_\pi\vec\pi, \ \vec v_\mu=\frac{g_\rho}{2}\vec\rho_\mu, \ \vec a'_\mu =\frac{g_\rho}{2}\vec a_{1\mu}.
\end{equation}
The renormalization constants $g_\sigma , g_\pi , g_\rho$ and masses of meson states can be expressed through the four independent parameters of the model: $\hat m$, $G_S$, $G_V$, and $\Lambda$
\begin{eqnarray}
\label{gs}
&&g_\sigma^2=\frac{4\pi^2}{N_cJ_1},\quad  g_\pi^2=Zg_\sigma^2,\quad  g_\rho^2=6g_\sigma^2, \\
\label{psm}
&&m_\pi^2=\frac{\hat mg_\pi^2}{mG_S},\quad m_\sigma^2=4m^2+Z^{-1}m_\pi^2, \\
\label{vamf}
&&m_\rho^2=\frac{6\pi^2}{N_cG_VJ_1},\quad m_{a_1}^2=m_\rho^2+6m^2,
\end{eqnarray}
where $Z=(1-2\kappa m^2)^{-1}$. From eqs. (\ref{vamf}), (\ref{kappa}) and (\ref{gs}) one obtains other useful relations
\begin{eqnarray}
&&m_{a_1}^2=Zm_\rho^2, \qquad (Z-1)m_\rho^2=6m^2, \\
&&m_\rho^2=\left(\frac{Z}{Z-1}\right)g_\rho^2 f_\pi^2.
\end{eqnarray}

From these expressions, one can see that a specific value, $Z=2$, corresponds to the celebrated Weinberg's result $m_{a_1}=\sqrt{2}m_\rho$ \cite{Weinberg67b,Weinberg69,Gilman68} and, at the same time, reproduces the KSRF formula for the $\rho$ coupling to the isospin current $m_\rho^2=2g_\rho^2f_\pi^2$ \cite{KS66,RF66}. Then the phenomenological value $m_\rho=775.26\pm 0.25\,\mbox{MeV}$ \cite{Patrignani16} gives us the estimates for the $a_1(1260)$ meson mass $m_{a_1}=1096\,\mbox{MeV}$, and for the constituent quark mass $m=m_\rho/\sqrt{6}=316\, \mbox{MeV}$.

The divergence of the axial-vector current, as it follows from (\ref{sb}), (\ref{ren}) and (\ref{psm}), is
\begin{equation}
\partial^\mu \vec J_\mu^A=-\frac{\partial\delta\mathcal L_B}{\partial\vec\beta}=\frac{\hat m g_\pi}{G_S}\vec \pi =\frac{m}{g_\pi}m_\pi^2\,\vec\pi.
\end{equation}
This relation is the standard partial conservation law of the axial current (PCAC relation). Consequently, we obtain the quark analog of the Goldberger-Treiman relation, which is $m=f_\pi g_\pi$, where $f_\pi=92$ MeV is the weak pion decay constant. Taking the value of $f_\pi$ as a third input to fix the parameters of the model, we determine $G_V$
\begin{equation}
G_V=\frac{1}{4f_\pi^2}\left(\frac{Z-1}{Z}\right).
\end{equation}
Again, for $Z=2$ we have $G_V=1.48\times 10^{-5}\,\mbox{MeV}^{-2}$. On the other hand, from the formula (\ref{vamf}) we find
\begin{equation}
J_1(m^2,\Lambda^2)=8\pi^2\left(\frac{f_\pi}{m_\rho}\right)^2\left(\frac{Z}{Z-1}\right).
\end{equation}
Solving this equation with respect to $\Lambda$, at $Z=2$, one obtains the value of the cutoff $\Lambda =1520\,\mbox{MeV}$.

Taking as the final input the value of the charged pion mass, $m_{\pi^\pm} =139.57\,\mbox{MeV}$, we are left with the system of two equations, (\ref{gap}) and (\ref{psm}), to find the values of the current quark mass $\hat m$, and the coupling $G_S$. They are
\begin{eqnarray}
&& G_S=\frac{\displaystyle m^2}{\displaystyle m_\pi^2f_\pi^2+\frac{3m^2}{2\pi^2}J_0} =3.28 \times 10^{-6}\,\mbox{MeV}, \\
&& \hat m=m\left(1-\frac{N_cG_S}{2\pi^2} J_0 \right)=1.7\,\mbox{MeV}.
\end{eqnarray}

We now consider the fundamental (for any model of hadrons) decay $\rho\to\pi\pi$. The corresponding vertex, as it follows from the above consideration, has the form
\begin{eqnarray}
\label{rpp}
{\cal L}_{\rho\pi\pi}=&-&i\frac{g_\rho}{4}\,\mbox{tr}\left(\rho_\mu [\pi , \partial^\mu\pi ] \right.\nonumber\\
&-&\left.\frac{Z-1}{2m_{a_1}^2}\,\rho_{\mu\nu}[\partial^\mu\pi , \partial^\nu\pi ]\right).
\end{eqnarray}
It is easily seen from this that on the $\rho\,$-meson mass shell the higher derivative term leads to the coupling of $\rho\pi\pi$ interactions
\begin{equation}
{\cal L}_{\rho\pi\pi}^{\rho -mass}=-i\frac{g_\rho}{4}\left(\frac{Z+1}{2Z}\right)\,\mbox{tr}(\rho_\mu [\pi , \partial^\mu\pi ]).
\end{equation}
Thus, as in the massive Yang-Millls approach treated in refs. \cite{Schwinger67,Schechter84a}, or in the hidden symmetry scheme \cite{Bando88}, the NJL model suffers from a problem of the strong momentum dependence of the $\rho\pi\pi$-vertex. The reason for this is that the higher derivative term in (\ref{rpp}) leads to a correction arising from the fact that the emitted pions are not soft. It reduces the ratio $g_{\rho\pi\pi}/g_\rho = (Z+1)/(2Z)$ from $1$ (soft pions) to the value $3/4$ at $Z=2$ (hard pions included).

Now,  we can deduce that the $\rho\to\pi\pi$ decay width $\Gamma_{\rho\pi\pi}=g_{\rho\pi\pi}^2(m_\rho^2-4m_\pi^2)^{\frac{3}{2}}/(48\pi m_\rho^2)$ is strongly suppressed $\Gamma_{\rho\pi\pi}=83.7\,\mbox{MeV}$, to be compared with the soft pions result $\Gamma_{\rho\pi\pi}=148.8\,\mbox{MeV}$ of the model if the higher derivative term in (\ref{rpp}) would be absent. The latter is in fair agreement with experiment $\Gamma_{\rho\pi\pi}^{exp}=149.1\pm 0.8\,\mbox{MeV}$ \cite{Patrignani16}. Most probably this shows that $\rho$ mesons couple universally to the isovector current $f_{\rho\pi\pi}=g_\rho$. However, we cannot neglect the gradient-coupling terms in (\ref{rpp}) because they are protected by chiral symmetry of the whole Lagrangian function (\ref{linv}). We also cannot simply add the higher derivative terms to the phenomenological meson Lagrangian, like it has been suggested with regard to this issue in other approaches \cite{Bando88b,Meissner88}. Those solutions are not applicable to the NJL model. The reason is very simple. The NJL Lagrangian is totally defined on the quark level. The bosonization procedure is uniquely determined and is not compatible with any ad hoc changes at the level of meson Lagrangian.

\section{New axial-vector field}
\label{NewField}
To resolve the outlined above problem, we suggest a simple modification of eq. (\ref{shift1}) by including the term with two derivatives
\begin{equation}
\label{shift2}
a_\mu = a''_\mu +\kappa m \partial_\mu p + i\bar\kappa [\tilde v_{\mu\nu}, \partial^\nu p],
\end{equation}
where $\tilde v_{\mu\nu}=\partial_\mu v_\nu -\partial_\nu v_\mu$, and $\bar\kappa$ is a constant to be fixed. The new term changes the transversal part of the axial-vector field, due to the hard pions effect.

This replacement alters the chiral transformation law of vector and axial-vector fields. Actually, this follows from the $SU(2)\times SU(2)$ invariance of the combination, which is bilinear in quark fields $q=(u,d)$
\begin{equation}
\label{start}
\delta [\bar q\gamma^\mu\left(v_\mu+\gamma_5 a_\mu\right) q]=0,
\end{equation}
where $\delta q=i(\alpha +\gamma_5\beta )q$ and $\delta \bar q=i\bar q (-\alpha +\gamma_5\beta )$.
Inserting (\ref{shift2}) in (\ref{start}) we obtain the following restriction imposed by chiral symmetry on spin-1 fields
\begin{equation}
\label{finish}
\delta \left[\bar q\gamma^\mu\left(v_\mu+\gamma_5 (a''_\mu +\kappa m \partial_\mu p + i\bar\kappa [\tilde v_{\mu\nu}, \partial^\nu p])\right) q\right]=0.
\end{equation}
Gathering separately the factors multiplying $\gamma^\mu$ and $\gamma^\mu\gamma_5$ we conclude that (\ref{finish}) is equivalent to
\begin{eqnarray}
\label{vnew}
\delta v_\mu &=& i[\alpha, v_\mu] +i[\beta, a''_\mu ] \nonumber \\
&+& i\kappa m[\beta, \partial_\mu p] -\bar\kappa [\beta, [\tilde v_{\mu\nu}, \partial^\nu p]], \\
\label{anew}
\delta a''_\mu &=& i[\alpha, a''_\mu] +i[\beta, v_\mu ] + \kappa m \{\beta, \partial_\mu s \} \nonumber \\
&+&\bar\kappa [[\beta, \tilde a''_{\mu\nu} +i\bar\kappa\Omega_{\mu\nu} ],\partial^\nu p] \nonumber \\
&+& i\bar\kappa [\tilde v_{\mu\nu}, \{\beta, \partial^\nu s\}],
\end{eqnarray}
where $\Omega_{\mu\nu}$ is a short-hand notation for
\begin{eqnarray}
\Omega_{\mu\nu}&=&\partial_\mu [\tilde v_{\nu\sigma}, \partial^\sigma p]-\partial_\nu [\tilde v_{\mu\sigma}, \partial^\sigma p] \\
&=&-[\partial_\sigma\tilde v_{\mu\nu}, \partial^\sigma p] +[\tilde v_{\nu\sigma},\partial_\mu\partial^\sigma p]-[\tilde v_{\mu\sigma},\partial_\nu\partial^\sigma p].\nonumber
\end{eqnarray}
This combination enters the chiral transformation law of the vector field strength
\begin{equation}
\delta\tilde v_{\mu\nu}=i[\alpha, \tilde v_{\mu\nu}] + i [\beta, \tilde a''_{\mu\nu}+i\bar\kappa\Omega_{\mu\nu}].
\end{equation}
One can see that the chiral partners $v_\mu$ and $a''_\mu$ transform now nonlinearly under chiral $SU(2)\times SU(2)$. This is not surprising, since a symmetry like chirality does not manifest itself in linear $\gamma_5$ invariance relations, but rather relates a set of processes protected by this symmetry \cite{Weinberg68}.

One must however verify that equations (\ref{vnew}) and (\ref{anew}) represent a self-consistent realization of $SU(2)\times SU(2)$. The consistency requirements that must be satisfied by the transformation rules are embodied in the Jacobi identities
 \begin{eqnarray}
 \label{jac1}
 \delta_{[12]}v_\mu &=& [\delta_1, \delta_2] v_\mu, \\
 \label{jac2}
 \delta_{[12]}a''_\mu &=& [\delta_1, \delta_2] a''_\mu.
 \end{eqnarray}
Let us check if they are fulfilled. By using the composition properties of infinitesimal parameters
\begin{eqnarray}
\label{latter}
&& i\alpha_{[12]}=[\alpha_1, \alpha_2] +[\beta_1, \beta_2], \nonumber \\
&& i\beta_{[12]}=[\alpha_1, \beta_2] +[\beta_1, \alpha_2],
\end{eqnarray}
we find after some algebraic calculations the desired result (\ref{jac1})
\begin{eqnarray}
 [\delta_1, \delta_2] v_\mu &=& i[\alpha_{[12]}, v_\mu ] +i[\beta_{[12]}, a''_\mu ] + i\kappa m [\beta_{[12]}, \partial_\mu p]\nonumber \\
 &-& \bar\kappa [\beta_{[12]}, [\tilde v_{\mu\nu},\partial_\nu p]] =\delta_{[12]}v_\mu.
 \end{eqnarray}
We leave it to the reader's pertinacity to show that the second Jacobi identity (\ref{jac2}) is also satisfied. Thus eqs. (\ref{vnew}) and (\ref{anew}) are admitted infinitesimal $SU(2)\times SU(2)$ chiral transformation laws.

The phenomenologically successful idea of the universality of the $\rho$-mesons \cite{Sakurai69} can be used to fix the coupling $\bar\kappa$. Its value controls the unwanted effect of hard pions in the $\rho\pi\pi$ vertex, and, in particular, can be chosen to suppress it.

Indeed, let us consider again the $\rho\pi\pi$ vertex. Due to the replacement (\ref{shift2}), it will get new contributions coming out of two terms of the Lagrangian density $\mathcal L_B$
\begin{eqnarray*}
&&\frac{1}{4G_V}\, \mbox{tr}\, a_\mu^2 \to -\frac{im\kappa\bar\kappa}{2G_V}\,\mbox{tr}\left(\tilde v_{\mu\nu}[\partial^\mu p,\partial^\nu p]\right), \nonumber \\
&& \frac{N_sJ_1}{16\pi^2}\,\mbox{tr}\,\left(\bigtriangledown_\mu p\right)^2 \to 4im\bar\kappa \frac{N_sJ_1}{16\pi^2}\,\mbox{tr}\left(\tilde v_{\mu\nu}[\partial^\mu p,\partial^\nu p]\right).
\end{eqnarray*}
Using eq. (\ref{kappa}), to sum these contributions, one finds that the vertex (\ref{rpp}) takes now the form
\begin{eqnarray}
\label{rpp2}
{\cal L'}_{\rho\pi\pi}&=&-i\frac{g_\rho}{4}\,\mbox{tr}\left(\rho_\mu [\pi , \partial^\mu\pi ] \right) \\
&+&i\frac{g_\rho}{2} (Z-1)\left(m\bar\kappa +\frac{\kappa}{12}\right)\mbox{tr}\left(\rho_{\mu\nu}[\partial^\mu\pi , \partial^\nu\pi ]\right). \nonumber
\end{eqnarray}
The second term represents the hard pions contribution, which leads, at $\bar\kappa=0$, to a strong momentum dependence of the $f_{\rho\pi\pi}(l^2)$ coupling
\begin{equation}
f_{\rho\pi\pi}(l^2)=g_\rho \left[1-2(Z-1)\left(m\bar\kappa +\frac{\kappa}{12}\right)l^2\right],
\end{equation}
where $l$ is the 4-momentum of the $\rho$ meson. As opposed to this, the problematic contribution to this coupling can be totally suppressed with the choice
\begin{equation}
\label{barkappa}
\bar\kappa =-\frac{\kappa}{12m}=-\frac{1}{24m^3}\left(\frac{Z-1}{Z}\right).
\end{equation}
In this case the model gives the perfect estimate for the decay width  $\Gamma_{\rho\pi\pi}=148.8\,\mbox{MeV}$, possessing universality even for on-shell $\rho$-mesons $l^2=m_\rho^2$, as it follows, for instance, in the dispersion theory, when only the $\rho$-meson pole saturates the isovector electromagnetic form factor of the hadronic matrix element $\langle A|j^i_\mu|A\rangle$.

It is of particular interest to apply our result to the $a_1\to\rho\pi$ decay. The amplitude which follows from the Lagrangian density $\mathcal L_B$ under the standard consideration ($\bar\kappa =0$) is
\begin{eqnarray}
\label{apr}
&& {\cal L}_{a_1\pi\rho}=\frac{i}{4}f_\pi g_\rho^2Z\,\mbox{tr}\left\{a_{1\mu}[\rho^\mu,\pi] \right. \nonumber \\
&&\left. +\frac{\kappa}{3}\left(\tilde\rho_{\mu\nu}[a_1^\mu ,\partial^\nu\pi]+\tilde a_{1\mu\nu}[\rho^{\mu},\partial^\nu\pi]\right)\right\},
\end{eqnarray}
and, correspondingly, it gives on the $a_1$ and $\rho$ mesons mass shell
\begin{equation}
{\cal L}_{a_1\pi\rho}^{a_1\!,\,\rho -mass}=\frac{i}{4}f_\pi g_\rho^2\,\mbox{tr}\,\rho_\mu[\pi, a_1^\mu ].
\end{equation}
Noting, that $f_\pi g_\rho =m_\rho\sqrt{(Z-1)/Z}$, we recognize here, at $Z=2$, the result obtained by Schwinger \cite{Schwinger67}.

Let us consider corrections induced by the terms with $\bar\kappa$. There are the following three new contributions
\begin{eqnarray*}
&&\frac{1}{4G_V}\, \mbox{tr}\, a_\mu^2 \to \frac{i\bar\kappa}{2G_V}\,\mbox{tr}\left( a''_{\mu}[\tilde v^{\mu\nu},\partial_\nu p]\right), \nonumber \\
&& \frac{N_sJ_1}{16\pi^2}\,\mbox{tr}\,\left(\bigtriangledown_\mu p\right)^2 \to i\bar\kappa m^2\frac{N_cJ_1}{2\pi^2}\,\mbox{tr}\left( a''_{\mu}[\tilde v^{\mu\nu},\partial_\nu p]\right), \nonumber \\
-&&\frac{N_cJ_1}{48\pi^2}\,\mbox{tr}\,a_{\mu\nu}^2\to -i\bar\kappa\frac{N_cJ_1}{24\pi^2}\,\mbox{tr}\,\left(\tilde a''_{\mu\nu}\Omega^{\mu\nu}\right).
\end{eqnarray*}
These can be summed into
\begin{equation}
\label{a1pr}
\Delta\mathcal L_{a_1\pi\rho}=i\bar\kappa\frac{N_cJ_1}{12\pi^2}\,\mbox{tr}\left(m_{a_1}^2 \! a''_{\mu}\, [\tilde v^{\mu\nu},\partial_\nu p]-\frac{1}{2}\tilde a''_{\mu\nu}\Omega^{\mu\nu}\right)
\end{equation}
After some reordering of derivatives in the second term (omitting total derivatives), we find
\begin{eqnarray}
\mbox{tr}\left(\tilde a''_{\mu\nu}\Omega^{\mu\nu}\right)&=&2\, \mbox{tr}\left(\tilde a''_{\mu\nu}\partial^\mu [\tilde v^{\nu\sigma}, \partial_\sigma p]\right)\nonumber \\
&=&-2\, \mbox{tr}\left(\partial^\mu \tilde a''_{\mu\nu} [\tilde v^{\nu\sigma}, \partial_\sigma p]\right)\nonumber \\
&\to&2m_{a_1}^2 \mbox{tr}\left(a''_{\nu}\, [\tilde v^{\nu\sigma}, \partial_\sigma p]\right).
\end{eqnarray}
Here, on the last stage, we show the result which one obtains for on-shell $a_1$-mesons. Inserted back in (\ref{a1pr}), this gives $\Delta\mathcal L_{a_1\pi\rho}=0$ for the physical $a_1$. Thus, the $\bar\kappa$ dependent part of the redefinition (\ref{shift2}) does not affect the decay width of $a_1\to\rho\pi$, yielding the old result of Schwinger $\Gamma_{a_1\to\pi\rho}\simeq 200\,\mbox{MeV}$ predicted by the phenomenological Yang-Mills Lagrangian.

Our result differs from the one of Schnitzer and Weinberg \cite{Weinberg67}. In the latter case the modifications of $\rho\pi\pi$ and $a_1\rho\pi$ vertices are expressed in terms of a parameter $\delta$. The value $\delta =-1$ leads to $\rho$-universality on the mass shell, and affects the $a_1\to\rho\pi$ amplitude, giving $\Gamma_{a_1\to\pi\rho}=60.8\,\mbox{MeV}$. Thus, there does not appear to be an obvious connection between our results and the results of \cite{Weinberg67}.

\section{Conclusions}
\label{concl}
The purpose of this paper has been to resolve the problem of the strong momentum dependence of the $\rho\pi\pi$-vertex. The aim was to find a general scheme which could be applicable to all known approaches including the NJL type models, where the solution had not been obtained untill now.

The problem has been resolved through an appropriate extension of the $\pi a_1$ diagonalization mechanism. The NJL model has been used as an example. Although we did not discuss in the text the applications of the method to other effective Lagrangians, the main idea is clearly presented and can be easily applied to any of the models reviewed, for instance, in \cite{Meissner88}.

The present results differ from the obtained by other approaches. More precise measurements of the $a_1$ decay widths are expected to help in the selection among this and other solutions.

\noindent {\it Acknowledgments:} I would like to acknowledge discussions with M. K. Volkov, S. B. Gerasimov and B. Hiller.

\end{document}